\documentstyle[twoside,fleqn,espcrc2,epsf]{article} 
\def\c{\mathop{\rm c}}
\def\cl{\mathop{\rm cl}}

\def\df{\mathop{\rm d}}
\def\dec{\mathop{\rm dec}}
\def\eq{\mathop{\rm eq}}
\def\etal{{\em et al}\,}
\def\F{\mathop{\rm F}}
\def\L{\mathop{\rm L}}
\def\max{\mathop{\rm max}}
\def\N{\mathop{\rm N}}

\def\pr{\mathop{\rm p}}
\def\R{\mathop{\rm R}}
\def\rec{\mathop{\rm rec}}

\def\with{\mathop{\rm with}}
\def\lesssim{\mathbin{\;\raise1pt\hbox{$<$}\kern-8pt\lower3pt\hbox{$\sim$}\;}}
\def\gtrsim{\mathbin{\;\raise1pt\hbox{$>$}\kern-8pt\lower3pt\hbox{$\sim$}\;}}
\def\95cl{\mathop{95 \%\ \rm c.l.}}
\newcommand{\case}[2]{{\textstyle\frac{#1}{#2}}}
\def\cm{\mathop{\rm cm}}
\def\dK{\mathop{\rm K}}
\def\eV{\mathop{\rm eV}}
\def\GeV{\mathop{\rm GeV}}
\def\gm{\mathop{\rm gm}}
\def\Gyr{\mathop{\rm Gyr}}
\def\keV{\mathop{\rm keV}}
\def\km{\mathop{\rm km}}
\def\kpc{\mathop{\rm kpc}}
\def\MeV{\mathop{\rm MeV}}
\def\Mpc{\mathop{\rm Mpc}}
\def\sec{\mathop{\rm sec}}
\def\H{\mathop{\rm H}}
\def\Htwo{\mathop{\rm D}}
\def\Hethree{^3{\rm He}}
\def\Hefour{^4{\rm He}}
\def\Liseven{^7{\rm Li}}
\title{Cosmological implications of neutrinos}
\author{Subir\ Sarkar \\ \bigskip \noindent 
        Theoretical Physics, University of Oxford,\\
        1 Keble Road, Oxford OX1 3NP, UK}
\begin{document}
\thispagestyle{empty}
\begin{abstract}
\begin{picture}(0,0)
\put(445,180){\makebox(0,0)[r]{OUTP-97-54P}}
\put(445,170){\makebox(0,0)[r]{hep-ph/9710273}}
\put(0,-500){\makebox(0,0)[l]{\sf (Plenary talk at the XVI International
Workshop on Weak Interactions and Neutrinos, Capri, June 22-28 1997)}}
\end{picture}
Massive neutrinos were the first proposed, and remain the most
natural, particle candidate for the dark matter. In the absence of
firm laboratory evidence for neutrino mass, considerations of the
formation of large scale structure in the universe provide a
sensitive, albeit indirect, probe of this possibility. Observations of
galaxy clustering and large angle anisotropy in the cosmic microwave
background have been interpreted as requiring that neutrinos provide
$\sim20\%$ of the critical density. However the need for such `hot'
dark matter is removed if the primordial spectrum of density
fluctuations is tilted below scale-invariance, as is often the case in
physically realistic inflationary models. This question will be
resolved by forthcoming precision measurements of microwave background
anisotropy on small angular scales. This data will also improve the
nucleosynthesis bound on the number of neutrino species and test
whether decays of relic neutrinos could have ionized the intergalactic
medium.
\end{abstract}
\maketitle
%
\section{A BIT OF HISTORY}

Several years before neutrinos had even been experimentally detected,
Alpher \etal \cite{afh53} noted that they would have been in thermal
equilibrium in the early universe ``\ldots through interactions with
mesons'' at temperatures above $5\MeV$. Below this temperature the
neutrinos ``\ldots freeze-in and continue to expand and cool
adiabatically as would a pure radiation gas''. These authors also
observed that the subsequent annihilation of $e^\pm$ pairs would heat
the photons but not the decoupled neutrinos so by entropy conservation
$T_\nu/T$ would decrease from its high temperature value of unity down
to $(4/11)^{1/3}$ at $T\ll\,m_e$.\footnote{Thus the present density of
massless relic neutrinos is
\begin{equation}
\label{nnu}
 \frac{n_\nu}{n_\gamma} = \left(\frac{T_\nu}{T}\right)^3
                          \frac{n_\nu}{n_\gamma}\mid_{T=T_{\dec}}
                      = \frac{4}{11} \left(\frac{3}{4}\frac{g_\nu}{2}\right) ,
\end{equation} 
where $g_\nu=2$ (left-handed neutrinos and right-handed antineutrinos)
and the factor 3/4 reflects Fermi versus Bose statistics. This would
also be true for massive neutrinos if $m_\nu\ll\,T_{\dec}$ so that the
neutrinos are relativistic at decoupling. Thus for a present blackbody
temperature $T_0=2.728\pm0.002\dK$ \cite{cobet0}, the abundance per
flavour is $(3/11)2\zeta(3)T_0^3/\pi^2\simeq112.3\cm^{-3}$. While
relativistic they retain a Fermi-Dirac distribution with phase-space
density
\begin{equation}
\label{fnu}
 f_{\nu} = \frac{g_\nu}{(2\pi)^3} 
           \left[\exp\left(\frac{p}{T_\nu}\right)+1\right]^{-1} .
\end{equation}
}

Subsequently, Chiu and Morrison \cite{cm61} calculated the rate for
$e^+e^-\rightleftharpoons\nu_e\bar{\nu_e}$ in a plasma to be
$\Gamma_{\nu}\approx\,G_{\F}^2T^5$ for the universal Fermi interaction
and Zel'dovich \cite{z6566} equated this to the Hubble expansion rate
in the radiation-dominated era,
\begin{equation}
\label{H}
 H = \sqrt{\frac{8\pi G_{\N} \rho}{3}} ,\quad {\with}\ 
     \rho = \frac{\pi^2}{30} g_* T^4
\end{equation}
(where $g_*$ counts the relativistic degrees of freedom), to obtain
the decoupling temperature $T_{\dec}(\nu_e)\simeq2\MeV$. (Neutral
currents were then unknown so $T_{\dec}(\nu_\mu)$ was estimated from
the reaction $\mu\rightleftharpoons\,e\bar{\nu_e}\nu_\mu$ to be
$12\MeV$. Later De Graaf \cite{dg70} noted that they would keep
$\nu_\mu$'s coupled to the plasma down to the same temperature as
$\nu_e$'s.\footnote{In fact $T_{\dec}(\nu_\mu,\nu_\tau)\simeq3.5\MeV$
while $T_{\dec}(\nu_e)\simeq2.3\MeV$ because of the additional charged
current reaction \cite{tdec}. Actually decoupling is not an
instantaneous process so the neutrinos are slightly heated by the
subsequent $e^+e^-$ annihilation, increasing the number density
(\ref{nnu}) by $\lesssim1\%$ \cite{boltz}.}) However Zel'dovich
\cite{z6566} and Chiu \cite{c66} concluded that relic neutrinos,
although nearly as numerous as the blackbody photons, cannot make an
interesting contribution to the cosmological energy density since they
are presumably massless.

Interestingly enough, some years earlier Pontecorvo and Smorodinski
\cite{ps61} had discussed the bounds set on the cosmological energy
density of MeV energy neutrinos (created e.g. by large-scale
matter-antimatter annihilation) using data from the Reines--Cowan and
Davis experiments. (They even suggested searching for GeV energy
neutrinos by looking for upward going muons in underground
experiments!) Not surprisingly these bounds were rather weak so these
authors stated somewhat prophetically that ``\ldots it is not possible
to exclude a priori the possibility that the neutrino and antineutrino
energy density in the Universe is comparable to or larger than the
average energy density contained in the proton rest mass''. Zel'dovich
and Smorodinski \cite{zs61} noted that better bounds can be set by the
limits on the total cosmological energy density
$\rho_0~(\equiv\Omega\rho_{\c}$) following from the observed present
expansion rate $H_0$ and age $t_0$ of the universe.\footnote{The
critical density is
$\rho_{\c}=3H_0^2/8\pi\,G_{\N}\simeq1.879\times10^{-29}h^2\gm\,\cm^{-3}$
where the Hubble parameter $h\equiv\,H_0/100\km\,\sec^{-1}\Mpc^{-1}$,
so $H_0^{-1}=9.778\,h^{-1}\Gyr$.} Of course they were still discussing
{\em massless} neutrinos. Weinberg \cite{w62} even speculated whether
a degenerate sea of relic neutrinos can saturate the cosmological
energy density bound and noted that such a sea may be detectable by
searching for (scattering) events beyond the end-point of the Kurie
plot in $\beta$-decay experiments!

Several years later, Gershte\u{\i}n and Zel'dovich \cite{gz66} made
the connection that if relic neutrinos are massive, then a bound on
the mass follows from requiring that $m_\nu\,n_\nu<\rho_0$. Using the
general relativistic constraint $\Omega\,t_0^2\,H_0^2<(\pi/2)^2$, they
derived $\rho_0<2\times10^{-28}\gm\cm^{-3}$ (just assuming
$t_0>5\Gyr$, i.e. that the universe is older than the Earth) and
inferred that $m_{\nu_e},m_{\nu_\mu}<400$~eV for a present photon
temperature of $3~\dK$. Unfortunately their calculation of the relic
neutrino abundance was erroneous. They took $g_\nu=4$, i.e. assumed
massive neutrinos to be Dirac particles with fully populated
right-handed (RH) states (although they acknowledged that according to
the $V-A$ theory such states are non-interacting and would thus not be
in equilibrium at $T_{\dec}$). Moreover they did not allow for the
decrease in the neutrino temperature relative to photons due to
$e^+e^-$ annihilation. Nevertheless their bound was competitive with
the best laboratory bound on $m_{\nu_e}$ and $10^4$ times better than
that on $m_{\nu_\mu}$ demonstrating the sensitivity (if not the
precision!) of cosmological arguments.

A better bound of $m_{\nu_\mu}<130$~eV was quoted by Marx and Szalay
\cite{ms72} who numerically integrated the cosmological Friedmann
equation from $\nu_\mu$ decoupling down to the present epoch, subject
to the condition $t_0>4.5\Gyr$. Independently Cowsik and McClleland
\cite{cm72a} used direct limits on $\Omega$ and $h$ to obtain
$m_\nu<8$~eV, assuming that $m_\nu=m_{\nu_e}=m_{\nu_\mu}$; however
they too assumed incorrectly that $T_\nu=T$ and that RH states were
fully populated. As Shapiro \etal \cite{stw80} first emphasized, even
if massive neutrinos are Dirac rather than Majorana, the RH states
have no gauge interactions so should have decoupled much earlier than
the left-handed ones. Thus subsequent entropy generation by massive
particle annihilations would have diluted their relic abundance to a
negligible level.\footnote{Although spin-flip scattering (at a rate
$\propto(m_\nu/T)^2$) {\em can} generate RH states, this can be
neglected for $m_\nu\ll1\MeV$. If RH neutrinos have new (superweak)
interactions, as in the
$SU(2)_{\L}\otimes\,SU(2)_{\R}\otimes\,U(1)_{B-L}$ model, then
corresponding bounds on their masses follow \cite{ot82}.} Now we
arrive \cite{bf81} at the modern version of the
`Gershte\u{\i}n-Zel'dovich bound' \cite{kt90}: the conservative limits
$t_0>10\Gyr$ and $h>0.4$ imply $\Omega\,h^2<1$
i.e. $\rho_0<10.54\keV\cm^{-3}$, so combining with eq.(\ref{nnu})
gives:\footnote{If the neutrinos are non-relativistic at decoupling,
then they drop out of chemical equilibrium with an abundance inversely
proportional to their self-annihilation cross-section so
$\Omega_{\nu}h^2\approx\,(m_\nu/2\GeV)^{-2}$ for $m_\nu\ll\,m_Z$
\cite{heavynu}. Thus neutrinos with a mass of ${\cal O}(\GeV)$ can
also account for the dark matter; however LEP has ruled out such (4th
generation) neutrinos. Conversely $\Omega_{\nu}h^2>1$ for the mass
range $\sim100\eV-2\GeV$, which is thus {\em forbidden} for any stable
neutrino having only electroweak interactions.}
\begin{equation}
\label{mnu}
 \Omega_\nu\,h^2 \simeq
 \sum_i \left(\frac{m_{\nu_i}}{94\eV}\right)\left(\frac{g_{\nu_i}}{2}\right)
 < 1 .
\end{equation} 
Note that $t_0=2/3H_0$ for a critical density universe so $t_0>10\Gyr$
requires $h\lesssim2/3$. For example a 30 eV neutrino would provide
$\Omega_\nu\simeq0.95$ (allowing $\Omega_{\N}=1-\Omega_\nu\simeq0.05$
in nucleons) if $h\simeq0.55$. According to a recent discussion
\cite{primrev}, most determinations \cite{h0} have converged on the
value $h=0.6\pm0.1$ corresponding to $t_0=9.3-13\Gyr$ for $\Omega=1$,
which is consistent with the recently revised age \cite{t0} of the
oldest stars in globular clusters. Measurements of the global
space-time geometry using Type I SN as `standard candles' are also
consistent with a critical density universe \cite{omega}, although
local dynamical measurements indicate a smaller value of
$\Omega\approx0.3$ \cite{p93}. This has led some cosmologists to
consider an open universe while others introduce a cosmological
constant ($\Lambda=1-\Omega\approx0.7$) to maintain a flat geometry,
notwithstanding the extreme fine-tuning of initial conditions implied
in either case.)

The above bound assumes conservatively that neutrinos constitute {\em
all} of the (dark) matter permitted by the global dynamics of the
universe. Further constraints must be satisfied if they are to cluster
on a specified scale (e.g. galactic halos or galaxy clusters) and
provide the dark matter whose presence is inferred from dynamical
measurements. Cowsik and McClleland \cite{cm72b} first suggested that
neutrinos with a mass of a few eV could naturally be the `missing
mass' in clusters of galaxies. This follows from the relation
$m_\nu^8\simeq1/G_{\N}^3r_{\cl}^3M_{\cl}$ (reflecting the Pauli
principle) which they obtained by modeling a cluster of mass
$M_{\cl}$ as a square potential well of core radius $r_{\cl}$ filled
with a Fermi-Dirac gas of neutrinos at zero temperature. Subsequently
Tremaine and Gunn \cite{tg79} noted that this provides a {\em lower}
bound on the neutrino mass. Although the microscopic phase-space
density (\ref{fnu}) is conserved for collisionless particles, the
`coarse-grained' phase-space density in bound objects can decrease
below its maximum value of $g_\nu/2(2\pi)^3$ during structure
formation. Modeling the bound system as an isothermal sphere with
velocity dispersion $\sigma$ and core radius
$r_{\cl}^2=9\sigma^2/4\pi\,G_{\N}\rho(r_{\cl})$ then gives
\begin{equation}
\label{tg}
 m_\nu > 120 \eV \left(\frac{\sigma}{100\,\km\,\sec^{-1}}\right)^{-1/4} 
         \left(\frac{r_{\cl}}{\kpc}\right)^{-1/2}.
\end{equation}
This is consistent with the cosmological upper bound (\ref{mnu}) down
to the scale of galaxies.\footnote{However there is a conflict for
smaller objects, viz. dwarf galaxies \cite{dg}. In fact the central
phase space density of observed dark matter cores in these structures
drops rapidly with increasing core radius, rather than being constant
as would be expected for neutrinos \cite{b97}.} Moreover since
neutrinos would cluster more efficiently in larger potential wells,
there should be a trend of increasing mass-to-light ratio with scale,
as seemed observationally to be the case \cite{ss81}.\footnote{This
too proved to be a problem later when it was recognized that the
actual increase is less than expected \cite{bfpr84}.}

\section{THE RISE AND FALL OF HOT DARK MATTER}

Such cosmological arguments became of particular interest in the
eighties after the ITEP tritium $\beta$-decay experiment claimed a
$\approx30$~eV mass for the electron neutrino. The attention of
cosmologists now focussed on how the large-scale structure (LSS) of
galaxies, clusters and superclusters \cite{lss} would have formed if
the universe is dominated by massive neutrinos. The basic picture is
that structure grows through gravitational instability from primordial
density perturbations \cite{struc}; these perturbations have now been
detected \cite{cobe} (see Figure~1) via the temperature fluctuations
they induce \cite{cmbth} in the cosmic microwave background (CMB). On
large scales ($\gtrsim30\Mpc$) the universe approaches spatial
homogeneity and gravitational dynamics is linear, while on smaller
scales structure formation is complicated by non-linear gravitational
clustering as well as non-gravitational (gas dynamic) processes. So
although we lack a standard model of galaxy formation \cite{gal}, the
physics of {\em large}-scale structure is sufficiently well understood
as to provide a reliable probe of the nature of the dark matter
\cite{p93}.

\begin{figure}[tbh]
\label{cobe4yr}
\center{\epsfysize\hsize\epsffile{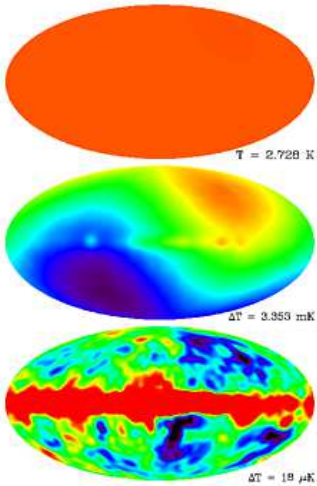}}
\caption{COBE \protect\cite{cobe} maps of the cosmic microwave
 background, showing the monopole, the dipole due to our motion
 relative to the frame in which the CMB is isotropic, and higher
 multipoles due to gravitational potential perturbations on the last
 scattering surface. The horizontal band is synchrotron emission from
 our Galaxy. (Courtesey of the COBE Science Working Group)}
\end{figure}

Density perturbations in a medium composed of relativistic
collisionless particles are subject to a form of Landau damping
(viz. phase-mixing through free streaming of particles from high to
low density regions) which effectively erases perturbations on scales
smaller than the free-streaming length
$\approx41\Mpc(m_\nu/30\eV)^{-1}$ \cite{hdm}. This is essentially the
(comoving) distance traversed by a neutrino from the big bang until it
becomes non-relativistic, and corresponds to the scale of
superclusters of galaxies. Thus huge neutrino condensations
(generically in the shape of `pancakes'), containing a mass
$\approx3\times10^{15}(m_\nu/30\eV)^{-2}M_\odot$, would have begun
growing at a redshift $z_{\eq}\approx7\times10^3(m_\nu/30\eV)$ when
the universe becomes matter-dominated and gravitational instability
sets in. This is well before (re)combination (at
$z_{\rec}\approx10^3$) so the baryons were still closely coupled to
the photons, while the neutrinos were mildly relativistic
($v/c\approx0.1$) hence `hot'. After the universe became neutral,
baryonic matter would have accreted into these potential wells,
forming a thin layer of gas in the central plane of the pancakes. Thus
superclusters would be the first objects to condense out of the Hubble
flow in a `hot dark matter' (HDM) cosmogony and smaller structures
such as galaxies would form later through their fragmentation.

The gross features of such a `top-down' model for structure formation
are compatible with several observed features of LSS, in particular
the distinctive `voids' and `filaments' seen in large galaxy
surveys. It was also noted that since primordial density perturbations
can begin growing earlier than in an purely baryonic universe, their
initial amplitude must have been smaller, consistent with extant
limits on the isotropy of the microwave background. Detailed studies
\cite{fs} found however that galaxies form late through the breakup of
the pancakes, at a redshift $z\lesssim1$, counter to observations of
galaxies, in particular quasars at $z>4$. (Another way of saying this
is that galaxies should have formed {\em last} in an HDM universe,
whereas our Galaxy is in fact dynamically much older than the local
group \cite{p84}.) There are other difficulties such as too large
`peculiar' (non-Hubble) velocities \cite{k83}, excessive X-ray
emission from baryons which accrete onto neutrino clusters
\cite{wdf84}, and too large voids \cite{zw90} (although detailed
simulations \cite{hdmok} showed later that some of these problems had
been exaggerated).

Therefore cosmologists soon abandoned HDM and turned, with
considerably more success, to cold dark matter (CDM) \cite{cdm},
i.e. particles which were non-relativistic at the epoch of
matter-domination. Detailed studies of CDM universes gave excellent
agreement with observations of galaxy clustering \cite{cdmrev} and
even led to progress in the understanding of galaxy formation
\cite{bfpr84,gal}. Thus a `standard CDM model' for large-scale
structure formation was established, viz. a critical density CDM
dominated universe with an initially scale-invariant spectrum of
density perturbations. Moreover particle physicists provided plausible
candidate particles, notably the neutralino in supersymmetric models
with conserved $R$-parity which naturally has a relic abundance of
order the critical density \cite{lsprev}.

\section{COBE AND THE ADVENT OF MIXED DARK MATTER}

Although the underlying physics is well known, cosmological structure
formation is a complex subject and some implicit assumptions must
necessarily be made in order to make progress. The key one concerns
the nature of the primordial density perturbations. Cosmologists
usually assume these to have a power spectrum of the scale-invariant
`Harrison-Zel'dovich' form:
\begin{equation}
 P (k) = \langle|\delta_k|^2\rangle = A k^n ,\quad \with\ n = 1 ,
\end{equation}
where $\delta_k\equiv\int\case{\delta\rho(\vec{x})}{\bar{\rho}}{\rm
e}^{-i\vec{k}\cdot\vec{x}}{\df}^3x$ is the Fourier transform of
spatial fluctuations in the density field (of wavelength
$\lambda=2\pi/k$). Moreover the perturbations are assumed to be
gaussian (i.e. different phases in the plane-wave expansion are
uncorrelated) and to be `adiabatic' (i.e. matter and radiation
fluctuate together). Concurrent with the above studies concerning the
nature of the dark matter, powerful support for this conjecture came
from the development of the `inflationary universe' model
\cite{kt90,infl}. Here the perturbations arise from quantum
fluctuations of a scalar field $\phi$, the vacuum energy of which
drives a period of accelerated expansion in the early universe. The
corresponding classical density perturbations have a spectrum
determined by the `inflaton' potential $V(\phi)$, with a power-law
index \cite{ll94}:
\begin{equation}
\label{nk}
 n (k) = 1 - 3 M^2 \left(\frac{V'}{V}\right)^2_\star
           + 2 M^2 \left(\frac{V''}{V}\right)_\star
\end{equation}
where $M\equiv(8\pi\,G_{\N})^{-1/2}\simeq2.4\times10^{18}\GeV$ is the
normalized Plank mass and $\star$ denotes that this is to be evaluated
when a mode of wavenumber $k$ crosses the `Hubble radius'
$H^{-1}$. Thus we see that for a sufficiently `flat potential' (as is
necessary to achieve sufficient e-folds of inflation to solve the
problems of the standard cosmology), the spectrum indeed has
$n\simeq1$.

As mentioned earlier, gravitational instability only sets in when the
universe becomes matter-dominated and this modifies the spectrum on
length scales smaller than the Hubble radius at this epoch, viz. for
$k>k_{\eq}^{-1}\simeq\,80h^{-1}\Mpc$. Thus the characteristics of the
dark matter can be encoded into a `transfer function' $T(k)$ which
modulates the primordial spectrum; for HDM this is an exponentially
dropping function while for CDM it is a more gradual power-law. Now
the power spectrum inferred from observations may be compared with
theoretical models but another problem arises concerning how we are to
normalize the amplitude of the primordial density
perturbations,\footnote{This is of course determined by the inflaton
potential $V(\phi)$ but we have as yet no definitive model of
inflation.} particularly since these are in the dark matter and may
differ significantly (i.e. be `biased') from the observable
fluctuations in the density of visible galaxies. Fortunately, the
primordial perturbations have another unique observational signature.
As mentioned earlier, they induce temperature fluctuations in the CMB
through the `Sachs-Wolfe effect' (gravitational red/blue shifts) on
large angular scales ($\gtrsim2^{0}$), corresponding to spatial scales
larger than the Hubble radius on the last scattering
surface.\footnote{Such fluctuations cannot have a causal origin in the
standard cosmological evolution and imply that the mechanism which
created them must have operated during a period of superluminal
expansion in the past, viz. inflation!} It was the COBE detection
\cite{cobe} of such fluctuations in 1992 that began the modern era of
cosmological structure formation studies.

\begin{figure}[tbh]
\label{wss_lss}
\epsfxsize\hsize\epsffile{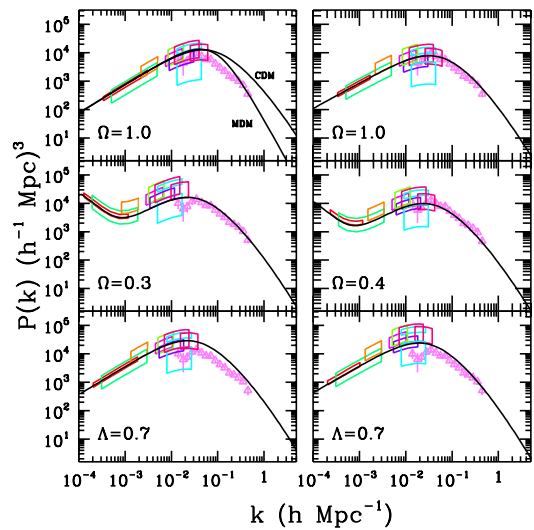}
\caption{The matter power spectrum as inferred from LSS and CMB data
 compared with theoretical models \protect\cite{wss95}. As seen top
 left, the excess small-scale power in the COBE-normalized standard
 CDM model ($n=1, \Omega_{\N}=0.03$, h=0.5) is reduced in the MDM
 model which has $\Omega_\nu=0.3$. This can also be achieved in a CDM
 model with a tilted spectrum ($n=0.9$) and higher nucleon density
 ($\Omega_{\N}=0.1, h=0.45$) as shown top right. The other panels show
 the expectations in an open universe (middle) and in a flat universe
 with a cosmological constant (bottom).}
\end{figure}

The quadrupole anisotropy in the CMB measured by COBE \cite{cobe}
allows a determination of the fluctuation amplitude at the scale,
$H_0^{-1}\simeq\,3000\,h^{-1}\Mpc$, corresponding roughly to the
present `size' of the universe. With this normalization it became
clear that a $\Omega_\nu\simeq1$ HDM universe indeed had too little
power on small-scales for adequate galaxy formation.\footnote{To save
HDM would require new sources of small-scale fluctuations, e.g. relic
topological defects \cite{hdmdef}, or {\em isocurvature} primordial
perturbations \cite{hdmiso}.} However it also became apparent (see
Figure~2) that the `standard CDM model' when normalized to COBE had
too {\em much} power on small-scales! It was thus a logical step to
invoke a suitable mixture of CDM and HDM to match the power spectrum
to the data on galaxy clustering and motions \cite{cobemdm}.

\begin{figure}[tbh]
\label{shafi_fig1}
\epsfxsize\hsize\epsffile{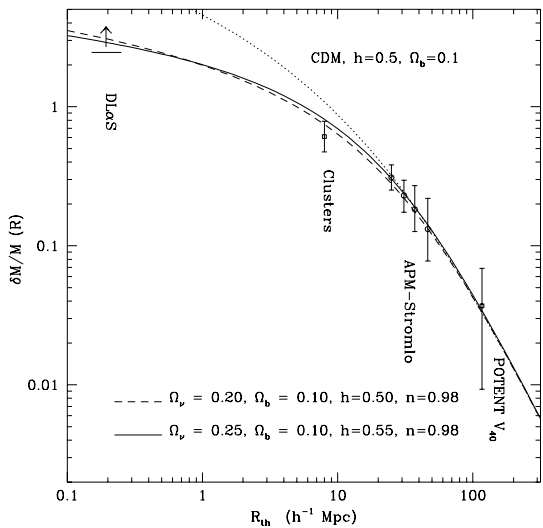}
\caption{The expected scale-dependence of mass fluctuations in
 supersymmetric inflationary MDM models compared with observations
 \protect\cite{shafi96}.}
\end{figure}

In fact the possibility that the dark matter may have both a hot and a
cold component had been discussed several years earlier, motivated by
theoretical considerations of SUSY GUTs, and the consequent advantages
for large-scale structure noted \cite{cphdm}. In the post-COBE era, a
number of detailed studies of mixed dark matter (MDM or CHDM)
universes have been performed and a neutrino fraction of about $20\%$
found to give the best match with observations \cite{mdm}. In Figure~3
we show a fit to observational data for a MDM model which also
incorporates a (globally) supersymmetric mechanism for inflation
\cite{shafi96}. The implied neutrino mass is about $5\eV$ and the
expectation (in the `see-saw' model for neutrino masses) is that this
is the $\nu_\tau$. Therefore CHORUS/NOMAD will provide a test of this
possibility (assuming that the CKM mixing with the lighter neutrinos
is of the same order as in the quark sector). More baroque schemes in
which two neutrinos have comparable masses may be constructed
(e.g. $m_{\nu_\mu}\approx\,m_{\nu_\tau}\approx2.5\eV$) if one wishes
to reconcile the LSND report of neutrino oscillations with the
indications of atmospheric neutrino oscillations \cite{LAlaw}. In this
case, both KARMEN and the forthcoming long-baseline experiments will
provide crucial tests.

However another way to reconcile a CDM universe with the small-scale
observations is to relax the underlying assumption that the primordial
spectrum is strictly scale-invariant. As shown in Figure~2, a mildly
`tilted' spectrum with index $n\approx0.9$ also gives a good fit to
the data \cite{tilt}. At first sight this might strike one as simply
introducing an additional parameter (although this is arguably no
worse than introducing an additional form of dark matter). However one
should really ask why the spectrum should be assumed to be {\em
exactly} scale-invariant in the first place!

\begin{figure}[tbh]
\label{tcdm_lss}
\epsfxsize\hsize\epsffile{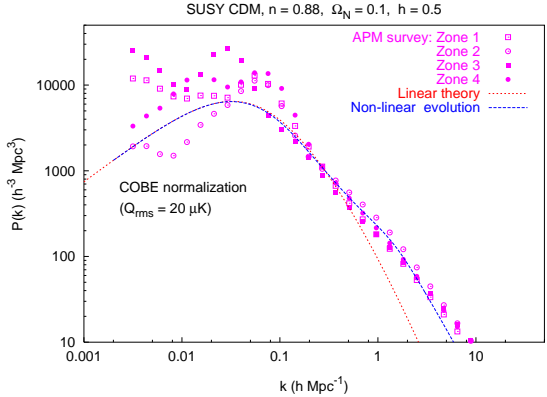}
\caption{Power spectrum of density perturbations in a supersymmetric
 inflationary CDM universe compared with APM survey data
 \protect\cite{susyinfl}.}
\end{figure}

As we saw earlier, the spectral index is determined by the slope and
curvature of the scalar potential at the epoch when the fluctuation on
a specified scale crosses the Hubble radius. The corresponding number
of e-folds before the end of inflation is just
\begin{equation}
\label{Nstar}
 N_\star(k) \simeq 51 + \ln\left(\frac{k^{-1}}{3000h^{-1}\Mpc}\right)  
\end{equation} 
(for typical choices of the inflationary scale, reheat temperature
etc). We see that fluctuations on the scales probed by LSS and CMB
observations ($\sim1-3000\Mpc$) are generated just $40-50$ e-folds
before the end of inflation. It would be not unnatural to expect the
inflaton potential to begin curving significantly as the end of
inflation is approached (e.g. in `new inflation' models). There are
certainly attractive models of inflation in which the spectrum is
significantly tilted in this region \cite{natural}. In a successful
inflationary model based on $N=1$ supergravity \cite{susyinfl}, the
spectral index is simply given by $n(k)\simeq(N_\star-2)/(N_\star+2)$
so is naturally $\approx0.9$ at these scales. In Figure~4 we compare
the power spectrum for such a tilted spectrum in a CDM universe (TCDM)
with data from the APM galaxy survey. It is seen that even the effects
of non-linear evolution (which generate a `shoulder' in the power
spectrum) at small scales can be successfully reproduced. Indeed even
MDM models \cite{mdm} now allow for the possibility that the
primordial spectrum may be tilted in order to achieve better fits to
the data. Figure~5 indicates schematically how the HDM fraction
required decreases as the tilt is increased \cite{dgt96}. There are
certainly differences in detail between MDM and TCDM models and new
observational constraints, e.g. the epoch of quasar formation or the
abundance of primordial Lyman-$\alpha$ clouds, may serve to
distinguish between them \cite{primrev}.\footnote{Such studies have
not however allowed for the possible scale-dependence of the tilt, as
is crucial in this context.} However a more powerful and unambiguous
discriminator is provided by the angular power spectrum of CMB
anisotropy.

\begin{figure}[tbh]
\label{omeganu-nk}
\epsfxsize\hsize\epsffile{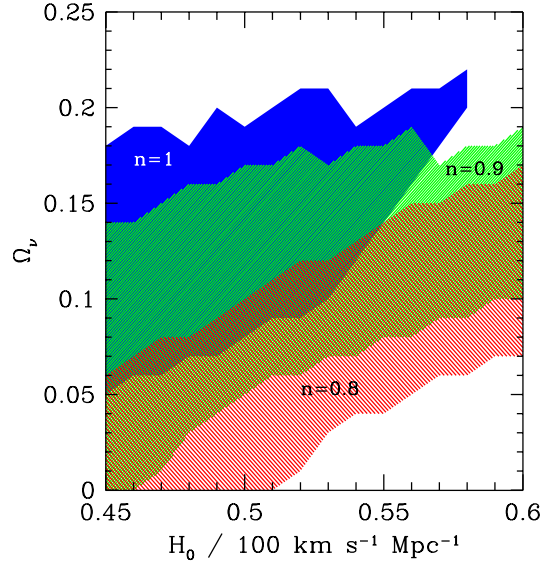}
\caption{Dependence of the HDM component required for successful LSS
 formation on the spectral index of the primordial perturbations
 \protect\cite{dgt96}.}
\end{figure}

In general a skymap of the CMB temperature can be decomposed into
spherical harmonics
\begin{equation}
 T (\theta, \phi) = \sum_{l=0}^{\infty} \sum_{m=-l}^{l}
                     a_{l}^{m} Y_{l}^{m} (\theta, \phi) ,
\end{equation} 
where the $l^{\rm th}$ multipole corresponds to an angle
$\approx200/\theta^{0}$ and probes spatial scales around
$k^{-1}\approx6000h^{-1}l^{-1}\Mpc$. In inflationary theories, the
fluctuations are gaussian so the co-efficients $a_{l}^{m}$ are
independent stochastic variables with zero mean and variance
$C_{l}=\langle|a_{l}^{m}|^2\rangle$; each $C_l$ has a $\chi^2$
distribution with $(2l+1)$ degrees of freedom \cite{cl}. For an
assumed set of cosmological parameters ($\Omega$, $H_0$,
$\Omega_{\N}$) and given the primordial density perturbation spectrum,
the $C_{l}$'s can be determined by solution of the Einstein-Boltzmann
equations which describe how the different components (photons, ions,
electrons, dark matter particles \ldots) evolve \cite{cmbfluc}. Thus
theoretical estimates of the power $(2l+1)C_{l}/2\pi$ at each
multipole can be compared with observations. The low multipoles (large
spatial scales) are sensitive to the primordial spectrum
alone,\footnote{In principle, primordial gravitational waves can also
make a contribution here but this is expected to be negligible in
realistic inflationary models \cite{l97}.} but the COBE measurement
(of the first $\sim$20 multipoles) finds $n=1.2\pm0.3$ \cite{cobe} so
cannot discriminate between a scale-invariant and a mildly tilted
spectrum. However forthcoming measurements with angular resolution far
superior to COBE's $\sim7^{0}$ will measure the power at higher
multipoles. The dominant features in the power spectrum here are the
`acoustic peaks', the most prominent at $l\approx200$, arising from
oscillations of the coupled plasma-photon fluids at last scattering
\cite{cmbth}.

\begin{figure}[tbh]
\label{mdm_cl}
\epsfxsize\hsize\epsffile{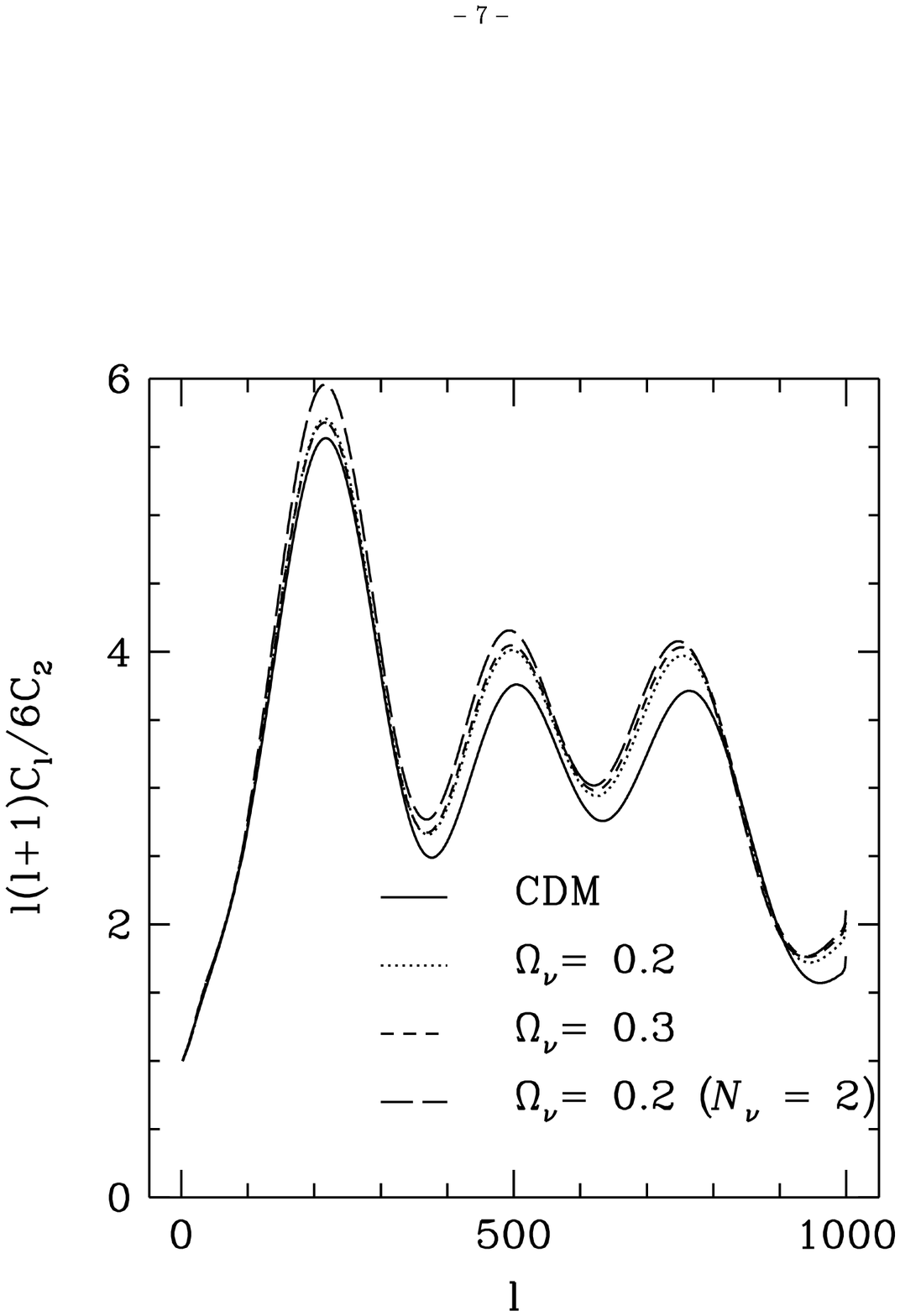}
\caption{Angular power spectrum of CMB anisotropy,assuming a
 scale-invariant spectrum, in CDM and MDM universes
 \protect\cite{dgs96}.}
\end{figure}

As seen in Figure~6, the expectations for CMB anisotropy in a MDM
universe do not differ significantly from a CDM universe having the
same initial perturbation spectrum. However if the primordial spectrum
is tilted, there is a significant suppression of the acoustic peaks at
high multipoles (see Figure~7). Whereas present observations of
small-angle anisotropy have a large scatter \cite{pdg}, forthcoming
measurements, in particular by the MAP \cite{map} and PLANCK
\cite{planck} satellite missions, will enable a definitive test. These
observations will also determine all the cosmological parameters
($\Omega$, $H_0$, $\Omega_{\N}$) to an accuracy of a few percent
\cite{cmball}, opening up a new era in cosmology.

\section{UNSTABLE NEUTRINOS?}

The cosmological bound (\ref{mnu}) is respected by the electron
neutrino for which the Particle Data Group \cite{pdg} now quotes
$m_{\nu_e}<15\eV$. Kinematic bounds on the masses of the other
neutrinos are much weaker ($m_{\nu_\mu}<170\keV$,
$m_{\nu_\tau}<24\MeV$) so in principle they may have masses in the
cosmologically forbidden range.

\begin{figure}[tbh]
\label{tcdm_cl}
\epsfxsize\hsize\epsffile{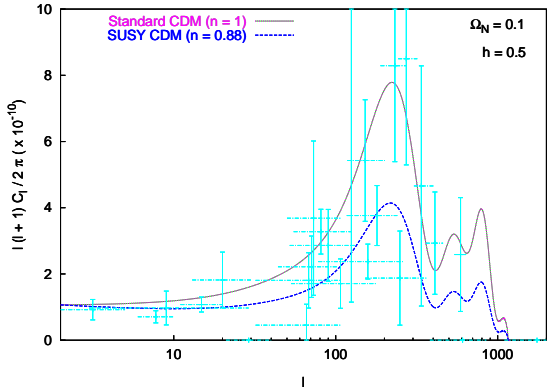}
\caption{Angular power spectrum of CMB anisotropy for a
 scale-invariant and a tilted spectrum \protect\cite{susyinfl}
 compared with observational data \protect\cite{pdg}.}
\end{figure}

There have been various suggestions (many motivated by the
now-withdrawn $17\keV$ neutrino discovery) that neutrinos may have new
interactions which enable them to decay or annihilate sufficiently
rapidly such that their relic abundance is reduced below the
cosmological limit. In general this can be ruled out if the decays (or
annihilations) create `visible' Standard Model particles,
e.g. $\nu'\to\nu\gamma$, $\nu'\to\nu_e\,e^+e^-$ \cite{gr95}. These
processes have not been seen in laboratory experiments \cite{of92}
(which probe short lifetimes), and would have affected cosmological
observables such as light element abundances \cite{s96} or the
radiation backgrounds \cite{kt90} (which are sensitive to long
lifetimes). The only possibility is to have such decays (or
annihilations) create hypothetical `invisible' particles,
e.g. Majorons (Goldstone bosons associated with lepton number
violation) \cite{v91}. These are fertile grounds for speculation, as
there are often no experimental constraints (by construction!) on such
hypotheses.

Many such proposals which {\em can} be experimentally tested have
already been falsified. For example, if tau neutrinos have a large
magnetic moment ($\sim10^{-6}\mu_{\rm B}$) their self-annihilations
are sufficiently boosted through $\gamma$ exchange that $\nu_\tau$'s
with mass of ${\cal O}(\MeV)$ may constitute the dark matter
\cite{g90}. However this was ruled out from the absence of anomalous
$\nu_\tau$ interactions in the BEBC beam dump experiment
\cite{wa66}. Another suggestion was that flavour-changing neutral
currents may allow the decay $\nu'\to\nu\nu\bar{\nu}$ to be
sufficiently fast \cite{nudec}. However the resultant breaking of the
GIM mechanism implies that the branching fraction of
$\nu'\to\nu\gamma$ is smaller only by a factor of ${\cal O}(\alpha)$
and this is observationally ruled out \cite{nurad}. Majoron models for
neutrino mass in which neutrinos may annihilate sufficiently rapidly
through Higgs exchange are ruled out by the LEP measurement of the
`invisible' width of the $Z^0$; only rather contrived singlet-Majoron
models survive and even these are severely constrained by the
non-observation of spectral features due to Majoron emission in
neutrinoless $\beta\beta$ decay \cite{gr95}. So although exotic decays
of e.g. the $\nu_\tau$ into singlet Majorons, cannot be definitively
excluded, it seems unlikely that it can thus evade the cosmological
bound and have a mass in the MeV range.

\begin{figure}[tbh]
\label{reion_cl}
\epsfxsize\hsize\epsffile{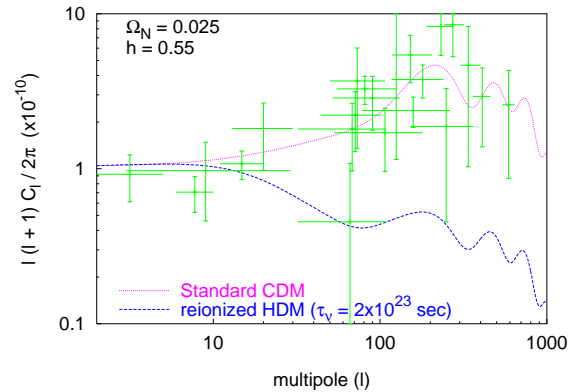}
\caption{Damping of the CMB anisotropy on small angular scales in an
 HDM universe reionized by radiative neutrino decays
 \protect\cite{jenni}.}
\end{figure}

For a neutrino mass subject to the cosmological bound (\ref{mnu}),
limits on the UV radiation background require the lifetime for
$\nu'\to\nu\gamma$ to be far longer than the age of the
universe. Apart from large-scale structure this is another
cosmological context where there may be an observational signal for
hot dark matter. Sciama \cite{decaynu} has argued that much of the
ionized hydrogen in both our Galaxy and in the intergalactic medium
cannot be accounted for in terms of conventional sources of UV photons
(of energy $\geq13.6\eV$), e.g. hot stars, supernovae or quasars. He
proposes that all such observations may be consistently understood if
the universe has $\Omega_\nu\simeq1$ in neutrinos of mass
$27.4\pm0.2\eV$ decaying radiatively with a lifetime of
$\sim2\pm1\times10^{23}\sec$. (Such a lifetime is smaller than
expected in most extensions of the Standard Model \cite{nudec} but
{\em can} arise in SUSY models with broken $R$-parity \cite{fastdec}.)
Again a decisive test of this theory is provided by CMB
observations. Following (re)combination the universe will soon be
reionized again due to the decaying neutrinos, thus washing out the
CMB anisotropy on small angular scales \cite{reion}. As seen in
Figure~8, the acoustic peaks in the power spectrum are thus severely
damped \cite{jenni}, a prediction that is already being tested by
ongoing experiments.

\section{THE BBN LIMIT ON $N_\nu$}

Hoyle and Taylor \cite{ht64} as well as Peebles \cite{p66} had
emphasized many years ago that new types of neutrinos (beyond the
$\nu_e$ and $\nu_\mu$ then known) would boost the relativistic energy
density hence the expansion rate (\ref{H}) during big bang
nucleosynthesis (BBN), thus increasing the yield of
$\Hefour$. Shvartsman \cite{s69} noted that new superweakly
interacting particles would have a similar effect. Subsequently this
argument was refined quantitatively by Steigman, Schramm and
collaborators \cite{chi}. In the pre-LEP era when the laboratory bound
on the number of neutrino species was not very restrictive
\cite{dss90}, the BBN constraint already indicated that at most one
new family was allowed \cite{bound}, albeit with rather uncertain
systematics \cite{eens86}. Although LEP now finds
$N_\nu=2.991\pm0.016$ \cite{pdg}, the cosmological bound is still
important since it is sensitive to {\em any} new light particle, not
just $SU(2)_{\L}$ doublet neutrinos, so is a particularly valuable
probe of new physics.  (The energy density of new light fermions $i$
is equivalent to an effective number
$\Delta\,N_{\nu}=\sum_{i}(g_i/2)(T_i/T_\nu)^4$ of additional doublet
neutrinos, where $T_i/T_\nu$ follows from considerations of their
(earlier) decoupling.)

The primordial mass fraction $Y_{\pr}(\Hefour)$ increases as
$\approx0.012\Delta\,N_\nu$ but it also increases logarithmically with
the nucleon density (usually parameterized as
$\eta\equiv\,n_{\N}/n_\gamma=2.728\times10^{-8}\Omega_{\N}h^2$). Thus
to obtain a bound on $N_\nu$ requires an upper limit on $Y_{\pr}$ {\em
and} a lower limit on $\eta$. The latter is poorly determined from
direct observations of luminous matter so must be derived from the
abundances of the other synthesized light elements, $\Htwo$,
$\Hethree$ and $\Liseven$, which are power-law functions of
$\eta$. The complication is that these abundances are substantially
altered in a non-trivial manner during the chemical evolution of the
galaxy, unlike $Y_{\pr}(\Hefour)$ which just increases by a few
percent due to stellar production. (This can be tagged via the
correlated abundance of oxygen and nitrogen which are made {\em only}
in stars.)

\begin{figure}[tbh]
\label{bbn}
\epsfxsize\hsize\epsffile{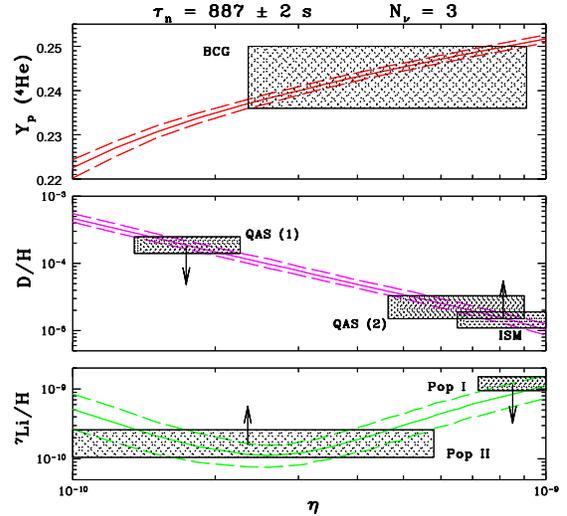}
\caption{Predicted light element abundances for the Standard Model
 versus the nucleon-to-photon ratio \protect\cite{us}. The $\95cl$
 limits determined by Monte Carlo reflect the uncertainties in input
 nuclear cross-sections and the neutron lifetime. Rectangles indicate
 observational determinations and associated `$\95cl$' bounds.}
\end{figure}

Even so, some cosmologists have used chemical evolution arguments to
limit the primordial abundances of $\Htwo$ and $\Hethree$ and thus
derived increasingly severe bounds on $N_\nu$ \cite{more}, culminating
in a recent one {\em below} 3 \cite{ohio}! However a more conservative
view \cite{us} is that there is no crisis with BBN if we recognize
that such arguments are rather dubious and consider only {\em direct}
measurements \cite{abund} of light element abundances, as shown in
Figure~9. The $\Hefour$ mass fraction is obtained from observations of
metal-poor blue compact galaxies by linear extrapolation to zero
nitrogen/oxygen abundance \cite{He4}; the upper limit is reliable, the
lower one less so. At present there are two conflicting measurements
of the $\Htwo$ abundance in quasar absorption systems
\cite{DQAShi,DQASlo}; the higher value \cite{DQAShi} is interpreted as
an upper limit. Also shown is the abundance in the interstellar medium
\cite{DISM} which provides a reliable lower limit. The $\Liseven$
abundance as measured in the hottest, most metal-poor halo stars
\cite{Li7PopII} as well as in disk stars \cite{Li7PopI} is shown and
interpreted as providing, respectively, reliable lower and upper
limits on its primordial value. Given these uncertainties, standard
BBN is consistent with observations for
$\eta\approx2-9\times10^{-10}$. Adopting the reliable limits,
$Y_{\pr}(\Hefour)<0.25$, $\Htwo/\H>1.1\times10^{-5}$ and
$\Liseven/\H<1.5\times10^{-9}$, and taking into account uncertainties
in nuclear cross-sections and the neutron lifetime by Monte Carlo, we
obtain \cite{us}
\begin{equation}
\label{Nnu4.53}
 N_{\nu}^{\max} = 3.75 + 78\ (Y_{\pr}^{\max} - 0.240) ,
\end{equation}
i.e. upto 1.5 additional (equivalent) neutrino species are allowed for
$\eta$ at its lowest allowed value. Other workers have applied
Bayesian likelihood methods to their adopted abundances (not limits as
above) to obtain $N_\nu<4-5$ \cite{them}. It is clear that the
restrictions on new physics are less severe than had been reported
previously \cite{more}.

\section{CONCLUSIONS}

Neutrino (hot) dark matter (with $\Omega_\nu\sim0.2$) is consistent
with but {\em not} required by our present understanding of
large-scale structure. Forthcoming CMB anisotropy measurements (in
particular by MAP and PLANCK) will resolve this question, as well as
provide precision determinations of cosmological parameters such as
$\Omega$ and $H_0$. The thermal history since (re)combination will
also be determined, enabling a test of the decaying neutrino theory.

Big bang nucleosynthesis permits at least one new neutrino, for
example a right-handed singlet neutrino which mixes with the
left-handed doublets. Again, CMB anisotropy observations will provide
an independent precise measurement of the nucleon density parameter
$\Omega_N$, so this bound will become a measurement.

As I have tried to indicate, the relationship between neutrinos and
cosmology has had a long history but with no definitive resolution as
yet. So far, intriguing hints from laboratory experiments have largely
driven the quest for cosmological consequences. With the renaissance
of observational cosmology, in particular studies of the cosmic
microwave background, the tables may well be turned in future.

\smallskip \noindent
{\bf Acknowledgments:} It is a pleasure to thank Vittorio Palladino
and Paolo Strolin for the invitation to participate in this enjoyable
meeting.


\begin{thebibliography}{99}

\bibitem{afh53}
 R. A. Alpher, J.W. Follin and R.C. Hermann, Phys. Rev. 92 (1953) 1347.

\bibitem{cobet0}
 D. Fixen \etal, Astrophys. J. 473 (1996) 576. 

\bibitem{cm61} 
 H.Y. Chiu and P. Morrison, Phys. Rev. Lett. 5 (1960) 573.

\bibitem{z6566} 
 Ya.B. Zel'dovich, Adv. Astron. Astrophys. 3 (1965) 241;
  Sov. Phys. Usp. 9 (1967) 602. 

\bibitem{dg70}
 T. De Graaf, Lett. Nuovo Cim. 4 (1970) 638. 

\bibitem{tdec}
 D. Dicus \etal, Phys. Rev. D26 (1982) 2694;
 K. Enqvist, K. Kainulainen and V. Semikoz, Nucl. Phys. B374 (1992) 392.

\bibitem{boltz}
 S. Dodelson and M.S. Turner, Phys. Rev. D46 (1992) 3372;
 S. Hannestad and J. Madsen, Phys. Rev. D52 (1995) 1764.

\bibitem{c66}
 H. Chiu, Ann. Rev. Nucl. Sci. 16 (1966) 591.

\bibitem{ps61}
 B. Pontecorvo and Ya. Smorodinski, Sov. Phys. JETP 14 (1962) 173.

\bibitem{zs61}
 Ya.B. Zel'dovich and Ya. Smorodinski, Sov. Phys. JETP 14 (1962) 647.

\bibitem{w62}
 S. Weinberg, Phys. Rev. 128 (1962) 1457; see also F. Vissani, hep-ph/9707343.

\bibitem{gz66}
 S.S. Gershte\u{\i}n and Ya.B. Zel'dovich, JETP Lett. 4 (1966) 120.

\bibitem{ms72}
 G. Marx and A.S. Szalay, Proc. Neutrino'72, Balatonf\"urd, Vol.1, p.123;
 A.S. Szalay and G. Marx, Astron. Astrophys. 49 (1976) 437.

\bibitem{cm72a} 
 R. Cowsik and J. McClleland, Phys. Rev. Lett. 29 (1972) 669.

\bibitem{stw80}
 S.L Shapiro, S.A. Teukolsky and I. Wasserman, Phys. Rev. Lett. 
  45 (1980) 669.
 
\bibitem{ot82}
 K.A. Olive and M.S. Turner, Phys. Rev. D25 (1982) 213.

\bibitem{bf81}
 J. Bernstein and G. Feinberg, Phys. Lett. 101B (1981) 39 
   (erratum 103B, 470).

\bibitem{kt90}
 E.W. Kolb and M.S. Turner, {\sl The Early Universe}
 (Addison-Wesley, 1990)

\bibitem{heavynu} 
 B.W. Lee and S. Weinberg, Phys. Rev. Lett. 39 (1977) 165;
 M. Vysotski\u{\i}, A. Dolgov and Ya.B. Zel'dovich, 
  JETP Lett. 26 (1977) 188. 

\bibitem{primrev}
 J.R. Primack, astro-ph/9707285.

\bibitem{h0}
 W. Freedman, astro-ph/9612024;
 A. Sandage and G. Tammann, astro-ph/9611170.

\bibitem{t0}
 B. Chaboyer \etal, astro-ph/9706128.

\bibitem{omega}
 S. Perlmutter \etal, astro-ph/9608192.

\bibitem{p93}
 P.J.E. Peebles, {\sl Principles of Physical Cosmology} (Princeton, 1993).

\bibitem{cm72b} 
 R. Cowsik and J. McClleland, Astrophys. J. 180 (1972) 7.

\bibitem{tg79}
 S. Tremaine and J. Gunn, Phys. Rev. Lett. 42 (1979) 407.

\bibitem{dg}
 e.g. D.N. Spergel, D.H. Weinberg and J.R. Gott, Phys. Rev. D38 (1988) 2014. 

\bibitem{b97}
 A. Burkert, astro-ph/9703057; 
 P. Salucci and A. Sinibaldi, Astr. Astrophys. 323 (1997) 1.

\bibitem{ss81}
 D.N. Schramm and G. Steigman, Astrophys. J. 243 (1981) 1

\bibitem{bfpr84}
 G.R. Blumenthal \etal, Nature 311 (1984) 517.

\bibitem{lss}
 G.~Efstathiou, in {\sl Cosmology and Large
  Scale Structure}, eds. R. Schaeffer \etal (Elsevier, 1996) p.133.

\bibitem{struc}
 E. Bertschinger, Physica D77 (1994) 354, in {\sl Cosmology and Large
  Scale Structure}, eds. R. Schaeffer \etal (Elsevier, 1996) p.273.

\bibitem{cobe}
 G. Smoot \etal, Astrophys. J. 396 (1992) L1; 
 C. Bennett \etal, Astrophys. J. 464 (1996) L1.

\bibitem{cmbth}
 R.K. Sachs and A.M. Wolfe, Astrophys. J. 147 (1967) 73; 
 W. Hu, N. Sugiyama and J. Silk, Nature 386 (1997) 37. 

\bibitem{gal}
 S.D.M. White, in {\sl Cosmology and Large Scale Structure}, 
  eds. R. Schaeffer \etal (Elsevier, 1996) p.349. 

\bibitem{hdm}
 A.G. Doroshkevich \etal, Sov. Astron. Lett. 6 (1980) 252;
 J.R. Bond, G. Efstathiou and J. Silk, Phys. Rev. Lett. 45 (1980) 1980.

\bibitem{fs}
 P.J.E. Peebles, Astrophys. J. 258 (1982) 415;
 A.A. Klypin and S.F. Shandarin, Mon. Not. R. Astr. Soc. 204 (1983) 891;
 C. Frenk, S.D.M. White and M. Davis, Astrophys. J. 271 (1983) 417;
 J.R. Bond and A. Szalay, Astrophys. J. 274 (1983) 443.

\bibitem{p84}
 P.J.E. Peebles, Science 224 (1984) 1385.
 
\bibitem{k83}
 N. Kaiser, Astrophys. J. 273 (1983) L17.

\bibitem{wdf84}
 S.D.M. White, M. Davis and C. Frenk, Mon. Not. R. Astr. Soc. 209 (1984) 27p.

\bibitem{zw90}
 Y. Zeng and S.D.M. White, Astrophys. J. 374 (1991) 1

\bibitem{hdmok}
 E. Braun, A. Dekel and P.R. Shapiro, Astrophys. J. 328 (1988) 34;
 J. Centrella \etal, Astrophys. J. 333 (1988) 333;
 R. Cen and J.P. Ostriker, Astrophys. J. 399 (1992) 331.

\bibitem{cdm}
 P.J.E. Peebles, Astrophys. J. 263 (1984) L1;
 J.R. Bond and G. Efstathiou, Astrophys. J. 285 (1984) L45.

\bibitem{cdmrev}
 C. Frenk, Phys. Scr. T36 (1991) 70;
 J.P Ostriker, Ann. Rev. Astron. Astrophys. 31 (1993) 689.

\bibitem{lsprev}
  G. Jungman, M. Kamionkowski and K. Griest, Phys. Rep. 267 (1996) 195.

\bibitem{infl}
 A.D. Linde, {\sl Particle Physics and Inflationary Cosmology}
  (Harwood Academic, 1990).

\bibitem{ll94}
 A.R. Liddle and D.H. Lyth, Phys. Rep. 231 (1993) 1.

\bibitem{wss95}
 D. Scott, J. Silk and M. White, Science 268 (1995) 829.

\bibitem{hdmdef}
 R. Brandenberger \etal, Phys. Rev. D59 (1987) 237;
 E. Bertschinger and P.N. Watts, Astrophys. J. 328 (1988) 23;
 J. Gratsias \etal, Astrophys. J. 405 (1993) 30;

\bibitem{hdmiso}
 P.J.E. Peebles, in {\sl The Origin and Evolution of Galaxies},
  ed. B.J.T. Jones (Reidel, 1983) p.143;
 N. Sugiyama, M. Sasaki and K. Tomita, Astrophys. J. 338 (1989) L45.

\bibitem{cobemdm}
 E.L Wright \etal, Astrophys. J. 396 (1992) L13; 
 M. Davis, F.J. Summers and D. Schegel, Nature 359 (1992) 393;
 A.N. Taylor and M. Rowan-Robinson, Nature 359 (1992) 396.

\bibitem{cphdm}
 Q. Shafi and F.W. Stecker, Phys. Rev. Lett. 53 (1984) 1292;
 R.K. Schaefer, Q. Shafi and F.W. Stecker, Astrophys. J. 347 (1989) 575;
 A. Van Dalen and R.K. Schaefer, Astrophys. J. 398 (1992) 33;

\bibitem{mdm}
 A. Klypin, \etal, Astrophys. J. 416 (1993) 1;
 Y.P. Jing \etal, Astron. Astrophys. 284 (1994) 703;
 C-P Ma and E. Bertshinger, Astrophys. J. 429 (1994) 22;
 D.Yu Pogosyan and A.A. Starobinsky, Astrophys. J. 447 (1995) 465;
 A. Liddle \etal, Mon. Not. R. Astr. Soc. 281 (1996) 531;
 A. Klypin, R. Nolthniius and J. Primack, Astrophys. J. 474 (1997) 533;
 C.C. Smith \etal, astro-ph/9702099.

\bibitem{shafi96}
 Q. Shafi and R.K. Schaefer, hep-ph/9612478.

\bibitem{LAlaw}
 J. Primack \etal, Phys. Rev. Lett. 74 (1995) 2160.

\bibitem{tilt}
 M. White {\em et al}, Mon. Not. R. Astr. Soc. 276 (1995) L69,
  283 (1996) 107.

\bibitem{natural}
 F.C. Adams {\em et al}, Phys. Rev. D47 (1993) 426.

\bibitem{susyinfl}
 G.G. Ross and S. Sarkar, Nucl. Phys. B461 (1996) 597; 
 J. Adams, G.G. Ross and S. Sarkar, Phys. Lett. B391 (1997) 271.

\bibitem{dgt96}
 S. Dodelson, E. Gates and M.S. Turner, Science 274 (1996) 69.

\bibitem{cl}
 P.J.E. Peebles, Astrophys. J. 263 (1982) L1;
 L. Abbott and M. Wise, Phys. Lett. 135B (1984) 279;
 M. White, D. Scott and J. Silk, Ann. Rev. Astron. Astrophys. 
  32 (1994) 329. 

\bibitem{cmbfluc}
 W. Hu \etal, Phys. Rev. D52 (1995) 5498;
 E. Bertschinger, astro-ph/9506070;
 U. Seljak and M. Zaldariagga, Ap. J. 469 (1996) 437.

\bibitem{l97}
D.H. Lyth, Phys. Rev. Lett. 78 (1997) 1861; hep-ph/9609431.

\bibitem{dgs96}
 S. Dodelson, E. Gates and A. Stebbins, Astrophys.J. 467 (1996) 10.

\bibitem{pdg}
 Particle Data Group, Phys. Rev. D54 (1996) 1, 
  update on http://pdg.lbl.gov/ .

\bibitem{map}
 http://map.gsfc.nasa.gov/ .

\bibitem{planck}
 http://astro.estec.esa.nl/SA-general/Projects/Cobras/cobras.html .

\bibitem{cmball}
 L. Knox, Phys. Rev. D52 (1995) 4307;
 G. Jungman \etal, Phys. Rev. D54 (1996) 1332; 
 M. Zaldarriaga, D. Spergel and U. Seljak, Astrophys. J. 488 (1997) 1;
 J.R. Bond, G. Efstathiou and M. Tegmark, astro-ph/9702100.

\bibitem{gr95}
 G. Gelmini and E. Roulet, Rep. Prog. Phys. 58 (1995) 1207.

\bibitem{of92}
 L. Oberauer and F. von Feilitzsch, Rep. Prog. Phys. 55 (1992) 1093.

\bibitem{s96}
 S. Sarkar, Rep. Prog. Phys. 59 (1996) 1493.

\bibitem{v91}
 J.W.F. Valle, Prog. Part. Nucl Phys. 26 (1991) 91; 
  these Proceedings.

\bibitem{g90}
 G.F. Giudice, Phys. Lett. B251 (1990) 460.

\bibitem{wa66}
 A.M. Cooper-Sarkar \etal, Phys. Lett. B280 (1992) 153.

\bibitem{nudec}
 A. De R\'ujula and S. Glashow, Phys. Rev. Lett. 45 (1980) 942;
 J. Maalampi and M. Roos, Phys. Rep. 186 (1990) 53. 

\bibitem{nurad}
 R.N. Mohapatra and P.B. Pal, {\sl Massive Neutrinos in Physics and
  Astrophysics}, (World Scientific, 1991).

\bibitem{decaynu}
 D.W. Sciama, {\sl Modern Cosmology and the Dark Matter Problem}
  (Cambridge, 1993), astro-ph/9703068, astro-ph/9709243.

\bibitem{fastdec}
 E. Roulet and D. Tommasini, Phys. Lett. 256B (1991) 218;
 F. Gabbiani, A. Masiero and D.W. Sciama, Phys. Lett. 259B (1991) 323.

\bibitem{reion}
 D. Scott, M.J. Rees and D.W. Sciama, Astron. Astrophys. 250 (1991) 295.

\bibitem{jenni}
 J.A. Adams, private communication. 

\bibitem{ht64}
 F. Hoyle and R.J. Tayler, Nature 203 (1964) 1108.

\bibitem{p66}
 P.J.E. Peebles, Phys. Rev. Lett. 16 (1966) 411;

\bibitem{s69}
 V.F. Shvartsman, JETP Lett. 9 (1969) 184.

\bibitem{chi}
 G. Steigman, D.N. Schramm and J. Gunn, Phys. Lett. 66B (1977) 202;
 G. Steigman, K.A. Olive and D.N. Schramm, Phys. Rev. Lett. 43 (1979) 239.

\bibitem{dss90}
 D. Denegri, B. Sadoulet and M. Spiro, Rev. Mod. Phys. 62 (1990) 1

\bibitem{bound}
 J. Yang \etal, Astrophys. J. 281 (1984) 493;  
 G. Steigman \etal, Phy. Lett. 176B (1986) 33.

\bibitem{eens86}
 J. Ellis \etal, Phy. Lett. 167B (1986) 457.

\bibitem{more}
 K.A. Olive \etal, Phy. Lett. B236 (1990) 454;
 T.P Walker \etal, Astrophys. J. 376 (1991) 51;
 C.J. Copi, D.N. Schramm and M.S. Turner, Science 267 (1995) 192.

\bibitem{ohio}
 N. Hata \etal, Phys. Rev. Lett. 75 (1995) 3977.

\bibitem{us}
 P.J. Kernan and S. Sarkar, Phys. Rev. D54 (1996) 3681;
 S. Sarkar, astro-ph/9611232.

\bibitem{abund}
 P. Molaro, astro-ph/9709245.

\bibitem{He4} 
 Y.I. Izotov, T.X. Thuan, and V.A. Lipovetsky,
  Astrophys. J. Suppl. 108 (1997) 1;
 B. Pagel \etal, Mon. Not. R. Astr. Soc. 255 (1992) 325;
 E. Skillman \etal, Ann. N.Y. Acad. Sci. 688 (1993) 739.

\bibitem{DQAShi}
 A. Songaila \etal, Nature 368 (1994) 599;
 R.F. Carswell \etal, Mon. Not. R. Astr. Soc. 268 (1994) L1;
 M. Rugers and C.J. Hogan, Astrophys. J. 459 (1996) L1, 
  Astron. J. 111 (1996) 2135;
 J.K. Webb \etal, Nature 388 (1997) 250.

\bibitem{DQASlo}
 D. Tytler, X-M. Fan, and S. Burles, Nature 381 (1996) 207;
 S. Burles and D. Tytler, astro-ph/9603070;
 D. Tytler, S. Burles and D. Kirkman, astro-ph/9612121.

\bibitem{DISM}
 P.R. McCullough, Astrophys. J. 390 (1992) 213;
 J.L. Linsky \etal, Astrophys. J. 402 (1993) 694, 451 (1995) 335.

\bibitem{Li7PopII}
 J.A.~Thorburn, Astrophys. J. 421 (1994) 318;
 P. Molaro, F. Primas and P. Bonifacio, Astron. Astrophys.  295 (1995) L47.

\bibitem{Li7PopI}
 L. Hobbs and C. Pilchowski, Astrophys. J. 334 (1988) 734;
 B. Chaboyer and P. Demarque, Astrophys. J. 433 (1994) 510. 

\bibitem{them}
 C.J. Copi, D.N. Schramm and M.S. Turner, Phys. Rev. D55 (1997) 3389;
 B.D. Fields \etal, New Astron. 1 (1996) 77;
 K.A. Olive and D. Thomas, Astropart. Phys. 7 (1997) 27.
 
\end{thebibliography}
\end{document}